\documentclass[amsmath,amssymb,twocolumn,prl,floatfix,showpacs,nofootinbib]{revtex4-1}
\usepackage{amsmath}
\usepackage{graphicx}
\usepackage{subfigure}
\usepackage{epstopdf}
\usepackage{color}
\usepackage{multirow}
\usepackage{setspace}
\usepackage{overpic}
\usepackage{amssymb}
\usepackage[colorlinks,urlcolor=blue, citecolor=blue,linkcolor=blue] {hyperref}
\usepackage{lineno}
\usepackage{bm}
\usepackage{rotating}
\usepackage[utf8]{inputenc}
\usepackage{float}
\usepackage{multirow}
\usepackage{lineno}
\usepackage{verbatim}
\usepackage{appendix}

\let\oldequation\equation
\let\oldendequation\endequation
\renewenvironment{equation}
 {\linenomathNonumbers\oldequation}
 {\oldendequation\endlinenomath}

\begin{document}

\title{\bf \boldmath
Observation of $J/\psi$ Electromagnetic Dalitz Decays to $X(1835)$, $X(2120)$ and $X(2370)$
}
\author{
M.~Ablikim$^{1}$, M.~N.~Achasov$^{10,c}$, P.~Adlarson$^{67}$, S. ~Ahmed$^{15}$, M.~Albrecht$^{4}$, R.~Aliberti$^{28}$, A.~Amoroso$^{66A,66C}$, M.~R.~An$^{32}$, Q.~An$^{63,49}$, X.~H.~Bai$^{57}$
% ...
\\
\vspace{0.2cm}
(BESIII Collaboration)\\
\vspace{0.2cm} {\it
$^{1}$ Institute of High Energy Physics, Beijing 100049, People's Republic of China\\
% ...
$^{4}$ Bochum Ruhr-University, D-44780 Bochum, Germany\\
% ...
$^{10}$ G.I. Budker Institute of Nuclear Physics SB RAS (BINP), Novosibirsk 630090, Russia\\
% ...
$^{15}$ Helmholtz Institute Mainz, Staudinger Weg 18, D-55099 Mainz, Germany\\
% ...
$^{28}$ Johannes Gutenberg University of Mainz, Johann-Joachim-Becher-Weg 45, D-55099 Mainz, Germany\\
% ...
$^{32}$ Liaoning Normal University, Dalian 116029, People's Republic of China\\
% ...
$^{49}$ State Key Laboratory of Particle Detection and Electronics, Beijing 100049, Hefei 230026, People's Republic of China\\
% ...
$^{57}$ University of Jinan, Jinan 250022, People's Republic of China\\
% ...
$^{63}$ University of Science and Technology of China, Hefei 230026, People's Republic of China\\
% ...
$^{66}$ University of Turin and INFN, (A)University of Turin, I-10125, Turin, Italy; (B)University of Eastern Piedmont, I-15121, Alessandria, Italy; (C)INFN, I-10125, Turin, Italy\\
$^{67}$ Uppsala University, Box 516, SE-75120 Uppsala, Sweden\\
% ...
$^{c}$ Also at the Novosibirsk State University, Novosibirsk, 630090, Russia\\
% ...
}
}

\begin{abstract}

Using a sample of about 10 billion $J/\psi$ events collected at a center-of-mass energy $\sqrt s = 3.097$ GeV with the BESIII detector, the electromagnetic Dalitz decays $J/\psi \to e^+e^- \pi^+ \pi^- \eta'$, with $\eta'\to\gamma\pi^+ \pi^-$ and $\eta'\to\pi^+\pi^-\eta$, have been studied. 
The decay $J/\psi \to e^+ e^- X(1835)$ is observed with a significance of $15\sigma$, and the transition form factor of $J/\psi\to e^+e^-X(1835)$ is presented for the first time. The intermediate states $X(2120)$ and $X(2370)$ are also observed in the $\pi^+ \pi^- \eta'$ invariant mass spectrum with significances of $5.3\sigma$ and $7.3\sigma$. The corresponding product branching fractions for $J/\psi\to e^+e^-X$, $X\to\pi^+\pi^-\eta'$ $(X=X(1835), X(2120)$ and $X(2370))$, are reported.  

\end{abstract}

\pacs{}

\maketitle

\oddsidemargin  -0.2cm
\evensidemargin -0.2cm 

The state $X(1835)$ was first discovered by the BESII experiment in $J/\psi \to \gamma \pi^+ \pi^- \eta'$ decays~\cite{gamx} in 2005.
In 2011, the discovery was confirmed by the BESIII experiment~\cite{X2120} in the same channel. 
Later, using an order of magnitude larger data sample, BESIII performed a study of the $\pi^+\pi^-\eta^{\prime}$ line shape of the $X(1835)$ and reported a significant abrupt change in the slope of the $\pi^+\pi^-\eta^{\prime}$ invariant mass 
distribution at the proton-antiproton ($p\bar{p}$) mass threshold~\cite{lineshape}. Further study indicates that the $X(1835)$ shares the same spin parity $0^{-+}$~\cite{kskseta, gamphi} with the $X(p\bar{p})$, a $p\bar{p}$ bound state, which has been observed by BES~\cite{ppbar} and confirmed by BESIII~\cite{ppbar2} and CLEO~\cite{cleo}. 
 Many theoretical speculations believe that the $X(1835)$ is exactly a kind of $p\bar{p}$ bound state~\cite{bound1,bound2,bound3}, however, the $X(1835)$ has not been observed in other processes, such as $\Upsilon(1 S) \rightarrow \gamma p \bar{p}$~\cite{upsilon}, $J / \psi \rightarrow \omega p \bar{p}$~\cite{omega} and $J / \psi \rightarrow \phi p \bar{p}$~\cite{phi}. The nature of the $X(1835)$ is still controversial. 
 
Furthermore, BESIII~\cite{X2120} observed two other resonance states, the $X(2120)$ and $X(2370)$, in $J/\psi \to \gamma \pi^+ \pi^- \eta'$ decays. The unprecedented sample of about 10 billion $J/\psi$ decays collected with the BESIII detector allows confirmation of the $X(1835)$, $X(2120)$ and $X(2370)$ states in different $J/\psi$ decay modes, in particular, 
$J/\psi\rightarrow e^+e^-\pi^+ \pi^- \eta'$.

These electromagnetic (EM) Dalitz decays, where an off-shell photon is internally converted into an $e^+e^-$ pair, 
provide an ideal opportunity to probe the structure of hadronic states and to investigate the fundamental mechanisms of 
the interactions between photons and hadrons~\cite{Landsberg}. 
Such EM decays are found to be simpler and they allow
one to make a more complete theoretical interpretation than it is the case for pure
hadronic interactions. Consequently, these EM Dalitz decays constitute a testing ground for
any theory describing the structure of strongly interacting particles~\cite{Landsberg}. 

In theory, for the EM Dalitz decay $J/\psi\rightarrow e^+e^-X(1835)$, assuming point-like particles, the branching
fraction of the Dalitz decay can be exactly described by quantum electrodynamics (QED) in the Standard Model~\cite{Landsberg},
and the theoretical branching ratio, $R\equiv\frac{\mathcal{B}\left(J / \psi \rightarrow e^{+} e^{-} X(1835)\right)}{\mathcal{B}(J / \psi \rightarrow \gamma X(1835)) }$, is calculated to be $9.80\times 10^{-3}$~\cite{ratio}, where many theoretical uncertainties cancel.
The first measurement of the branching fraction of the EM Dalitz decay and the ratio $R$ will provide a straightforward test of this point-like QED prediction.

 The EM Dalitz decay $J/\psi\rightarrow e^+e^-X(1835)$ gives access to the EM transition form factors (TFFs) between the $J/\psi$ and $X(1835)$ states. The $q^2$-dependent differential decay rate of 
$J/\psi\rightarrow e^+e^-X(1835)$ normalized to the corresponding radiative decay $J/\psi\rightarrow \gamma X(1835)$ 
can be expressed as~\cite{Landsberg,DIY,formfactor}
\begin{equation}\label{eq0}
\begin{aligned}
\frac{d \Gamma\left(J/\psi \rightarrow X(1835) e^{+} e^{-}\right)}{d q^{2} \Gamma(J/\psi  \rightarrow X(1835) \gamma)}
=&\left|F\left(q^{2}\right)\right|^{2} \times\left[\operatorname{QED}\left(q^{2}\right)\right]
\end{aligned}
\end{equation}
where the normalized TFF for the $J/\psi\rightarrow X(1835)$ transition is defined as $\left|F\left(q^{2}\right)\right|^{2}$, $q^2$ is the square of the invariant mass of the $e^+e^-$ pair,
and $\left[\operatorname{QED}\left(q^{2}\right)\right]$~\cite{formfactor} represents the QED ratio calculated for point-like particles.  Experimentally, 
the TFF is directly accessible by comparing the measured invariant mass spectrum of the lepton
pairs from the Dalitz decays with the point-like QED prediction. The $q^2$-dependent TFF, 
which reflects the deviation from the point-like particle assumption~\cite{pointlike}, can provide additional information on the interactions between
$J/\psi$ and $X(1835)$, and serves as a sensitive probe into their internal structure. 
Furthermore, it can possibly help to distinguish the transition mechanisms based on the $q\bar{q}$ scenario 
and other solutions which alter the simple quark model picture. To be specific, besides the $p\bar{p}$ bound state, there are many different theoretical models for the nature of the $X(1835)$, such as a pseudoscalar glueball \cite{glueball1,glueball2, glueball3}, an excited $\eta'$ state \cite{etaprime} or an excited $\phi$ state \cite{phiprime}, etc. It could be most likely a mixing state of $p \bar{p}$ and $s \bar{s}$ and this idea has been investigated in e.g., Ref.~\cite{ppss}. Different theoretical models result in different TFFs. Our measurement of the $q^2$-dependent TFF will be helpful for the theorists when they proceed the predictions based on different assumptions of the $X(1835)$ structure.

In this Letter, we report the observation of the EM Dalitz decays $J / \psi \rightarrow e^{+} e^{-} X$, where $X$ represents the $X(1835)$, $X(2120)$ and $X(2370)$, and the measurement of TFF for the  
$J / \psi \rightarrow e^{+} e^{-} X(1835)$ transition. 

This work is performed with a sample of $(1.0087\pm0.0044)\times10^{10}$ $J/\psi$ events collected at the center-of-mass energy 3.097 GeV with the BESIII detector operating at the Beijing Electron Positron Collider (BEPCII)~\cite{Yu:IPAC2016-TUYA01}.
The total number of $J / \psi$ events collected in the years of 2009, 2012, 2018, and 2019 is determined using inclusive $J / \psi$ decays with the method described in Ref.~\cite{njpsi2017}. 
The BESIII detector is composed of a helium-based main drift chamber (MDC), a time-of-flight system (TOF), a CsI(Tl) electromagnetic calorimeter (EMC), and a muon counter (MUC).   
Details about the design and performance of the BESIII detector are given in Refs.~\cite{BESIII,whitepaper}.
Monte Carlo (MC) simulated data samples are produced with a {\sc Geant4}-based~\cite{geant4} software package, which includes the geometric description of the BESIII detector and the detector response. An inclusive MC sample containing about 10 billion $J/\psi$ events which includes both the production of the $J/\psi$ meson and the continuum processes incorporated in the event generator {\sc kkmc}~\cite{kkmc} is used to study possible physics backgrounds. The known decay modes of the $J/\psi$ meson are modeled with the event generator {\sc evtgen}~\cite{evtgen} using branching fractions taken from the Particle Data Group (PDG)~\cite{pdg2020}, and the remaining unknown charmonium decays are modeled with the event generator {\sc lundcharm}~\cite{lundcharm}. Final state radiation from charged final state particles is
incorporated using the PHOTOS package~\cite{PHOTOS}.
For the simulation of the signal processes, the three $J/\psi\to e^+e^-X$ decay modes are generated with angular distributions according to the amplitude squared in Eq.~(3) of Ref.~\cite{DIY} taking into account the $J/\psi$ polarization effect. The final state $\eta'$ candidates from $X\to\pi^+\pi^-\eta'$  are reconstructed with two decay modes: $\eta'\to\gamma\pi^+\pi^-$ and $\eta'\to\pi^+\pi^-\eta,~\eta\to\gamma\gamma$.

Charged tracks are reconstructed in the MDC with a polar-angle ($\theta$) range of $|\rm{cos}\theta|<0.93$.  The distance of closest approach of each track to the interaction point (IP) must be less than 10 cm along the $z$-axis, and less than 1 cm in the transverse plane. The number of charged tracks is required to be six, with a net charge equal to zero, and exactly two tracks are identified as the electron-positron pair in the final state. 
Particle identification (PID) for charged tracks combines measurements of the energy deposited in the MDC ($dE/dx$), the 
flight time in the TOF and the EMC information to form 
probabilities $Prob(i)_{i = e; \pi; K}$.
A track is considered an electron when $Prob(e)$ is larger than $Prob(\pi)$ and $Prob(K)$, and the remaining candidates are identified as pions. 
Photon candidates are required to have a minimum energy of 25 MeV in the EMC barrel region ($| \cos \theta| < 0.80$) or 50 MeV in the end-cap region ($0.86 < | \cos \theta| < 0.92$), and to be separated from charged tracks by more than 10$^\circ$. 

\begin{figure}[htp]
  \centering
\includegraphics[width=1\linewidth]{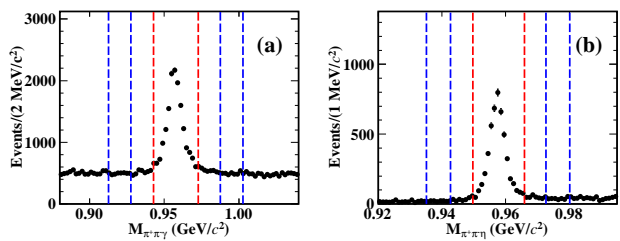}
  \caption{\small
Invariant mass distributions for selected (a) $\eta'\to\gamma\pi^+\pi^-$ and (b) $\eta'\to\pi^+\pi^-\eta$ candidates. The dots with error bars are data, the red vertical dashed lines indicate the $\eta'$ mass windows, and the pairs of blue dashed lines (left and right of the signal peak) indicate the $\eta'$ sideband regions.
}
\label{fig:comb4fig}
\end{figure}

For the $\eta'\to\gamma\pi^+\pi^-$ decay mode, a four-constraint (4C) kinematic fit imposing conservation of the initial energy and momentum is performed to the hypothesis of $e^+e^-\pi^+\pi^-\pi^+\pi^-\gamma$. For the events with more than one photon candidate, the combination with the minimum $\chi^2_{4\rm{C}}$  is selected, and events 
with $\chi^2_{4\rm{C}} < 60$ are retained. The $\chi^2$ selection requirement is optimized for maximal signal ($S$) to background ($B$) ratio, i.e. $S/\sqrt{S+B}$.
In addition, events with $M_{\gamma e^+ e^-} < 210$~MeV/$c^2$, $490 < M_{\gamma e^+ e^-} < 600$~MeV/$c^2$ and $700< M_{\gamma e^+ e^-} < 820$~MeV/$c^2$ are removed to suppress background events from $J/\psi\to\pi^0\pi^+\pi^-\pi^+\pi^-$, $J/\psi\to\eta\pi^+\pi^-\pi^+\pi^-$, $J/\psi\to\omega\pi^+\pi^-\pi^+\pi^-$ ($\pi^0, \eta \to\gamma e^+e^-$, $\omega\to \pi^0\gamma, \pi^0 e^+e^-$), respectively. Here, all mass windows correspond to about three times the invariant-mass resolution. 
Since the $\pi^+\pi^-$ pairs from $\eta' \to \gamma\pi^+\pi^-$ are dominantly from the $\rho^0$ resonance, their invariant mass must be in the range $575 < M_{\pi^+\pi^-} < 920$~MeV/$c^2$.
As shown in Fig~\ref{fig:comb4fig}(a), $\eta^{\prime}$ candidates are selected with $|M_{\gamma\pi^+\pi^-} - m_{\eta'}| < 15$~MeV/$c^2$. 
Events with $|M_{e^+e^-} - m_{\omega}| < 25$~MeV/$c^2$ and $|M_{e^+e^-} - m_{\phi}| < 30$~MeV/$c^2$ are removed to suppress background events from $J/\psi \to \omega\pi^+\pi^-\eta'~(\omega\to e^+e^-)$ and $J/\psi \to \phi\pi^+\pi^-\eta'~(\phi\to e^+e^-)$, respectively. The nominal masses of $\eta',~\omega,~\phi$ are taken from the PDG~\cite{pdg2020}.  Photons with energy larger than 1.02 MeV may convert to $e^+e^-$ pairs, and this usually occurs in the beam pipe and inner wall of the MDC, leading to a $\gamma$ conversion background. To reject this background from $J/\psi\to \gamma \pi^+\pi^-\eta'$, the candidate events must satisfy $R_{xy} < 2$~cm, where $R_{xy}$ is the distance from the reconstructed vertex of the $e^+e^-$ pair to the IP in the $x-y$ plane based on a $\gamma$ conversion finder algorithm \cite{gamconvfinder}. With the requirements above, the remaining yields for the $J/\psi\to\gamma X(1835), \gamma X(2120), \gamma X(2370)$ decays in the MC simulations of the $\gamma$ conversion background are $32\pm5,~6\pm4,~7\pm8$, respectively. These yields are subtracted from the corresponding number of signal events in the fit to data; the same treatment is used for the $\eta'\to\pi^+\pi^-\eta$ decay mode. 

For the $\eta'\to\pi^+\pi^-\eta$ decay mode, a five-constraint (5C) kinematic fit imposing conservation of the initial energy and momentum is performed to the $e^+e^-\pi^+\pi^-\pi^+\pi^-\eta$ hypothesis, and the invariant mass of $\gamma\gamma$ candidates is constrained to the known mass of the $\eta$ meson~\cite{pdg2020}. For events with more than two photon candidates, the combination with the smallest $\chi^2_{5\rm{C}}$ is selected, and the events with $\chi^2_{5\rm{C}} < 60$ are retained.
The $\eta'$ candidate is reconstructed from $\pi^+\pi^-\eta$ combinations in an event, and all combinations are kept. As shown in Fig~\ref{fig:comb4fig}(b), $\eta'$ candidates are selected with $|M_{\pi^+\pi^-\eta} - m_{\eta'}| < 8.1$ MeV/$c^2$. The requirements on $R_{xy}$ and $M_{e^+e^-}$ are the same as for the $\eta'\to\gamma\pi^+\pi^-$ decay mode. In total, $20\pm3,~3\pm2,~3\pm3$ events are selected for the $J/\psi\to\gamma X(1835),~\gamma X(2120),~\gamma X(2370)$ decay modes in the MC simulations of the $\gamma$ conversion background.   

\begin{figure*}
  \centering
\includegraphics[width=0.45\linewidth]{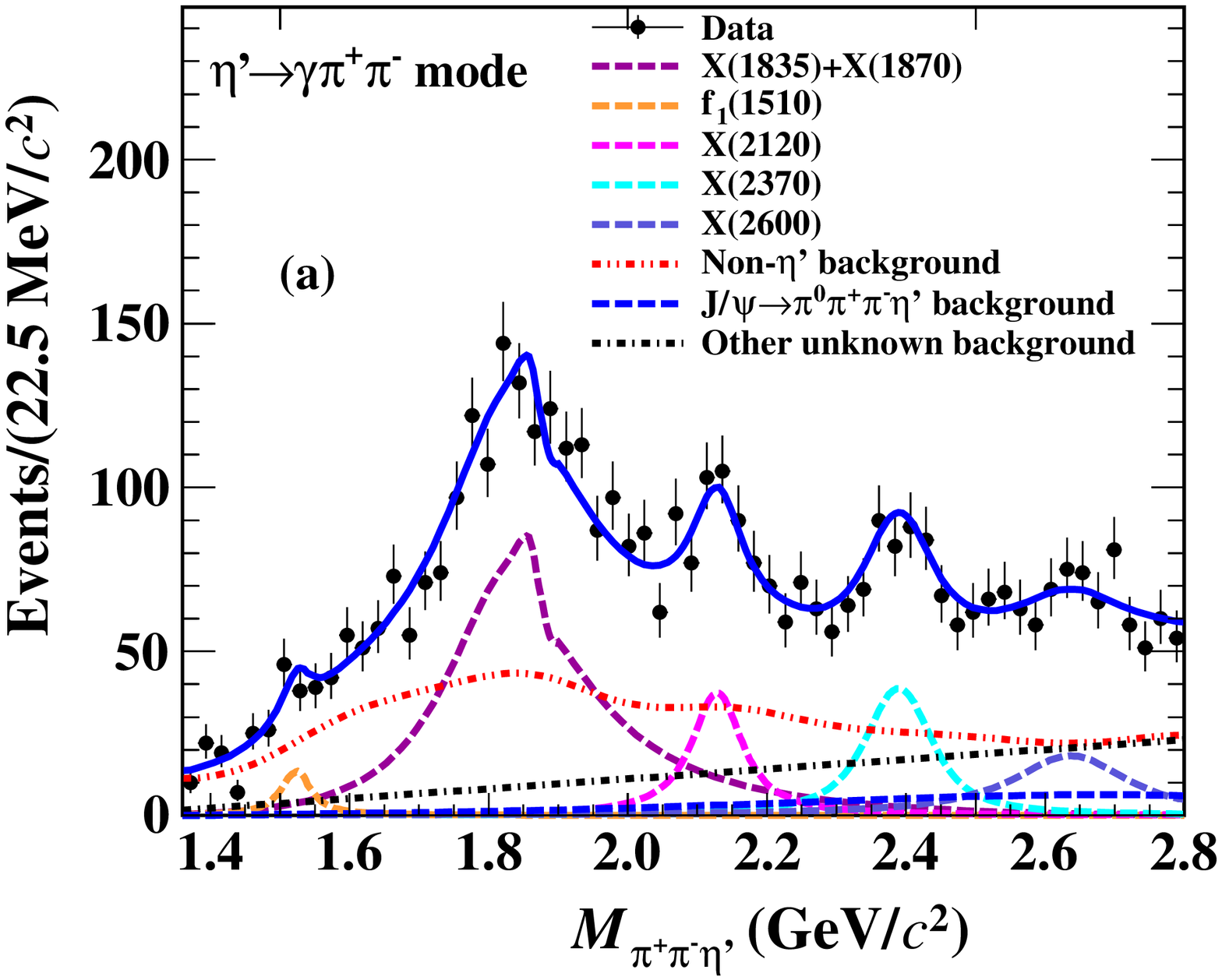}
\includegraphics[width=0.45\linewidth]{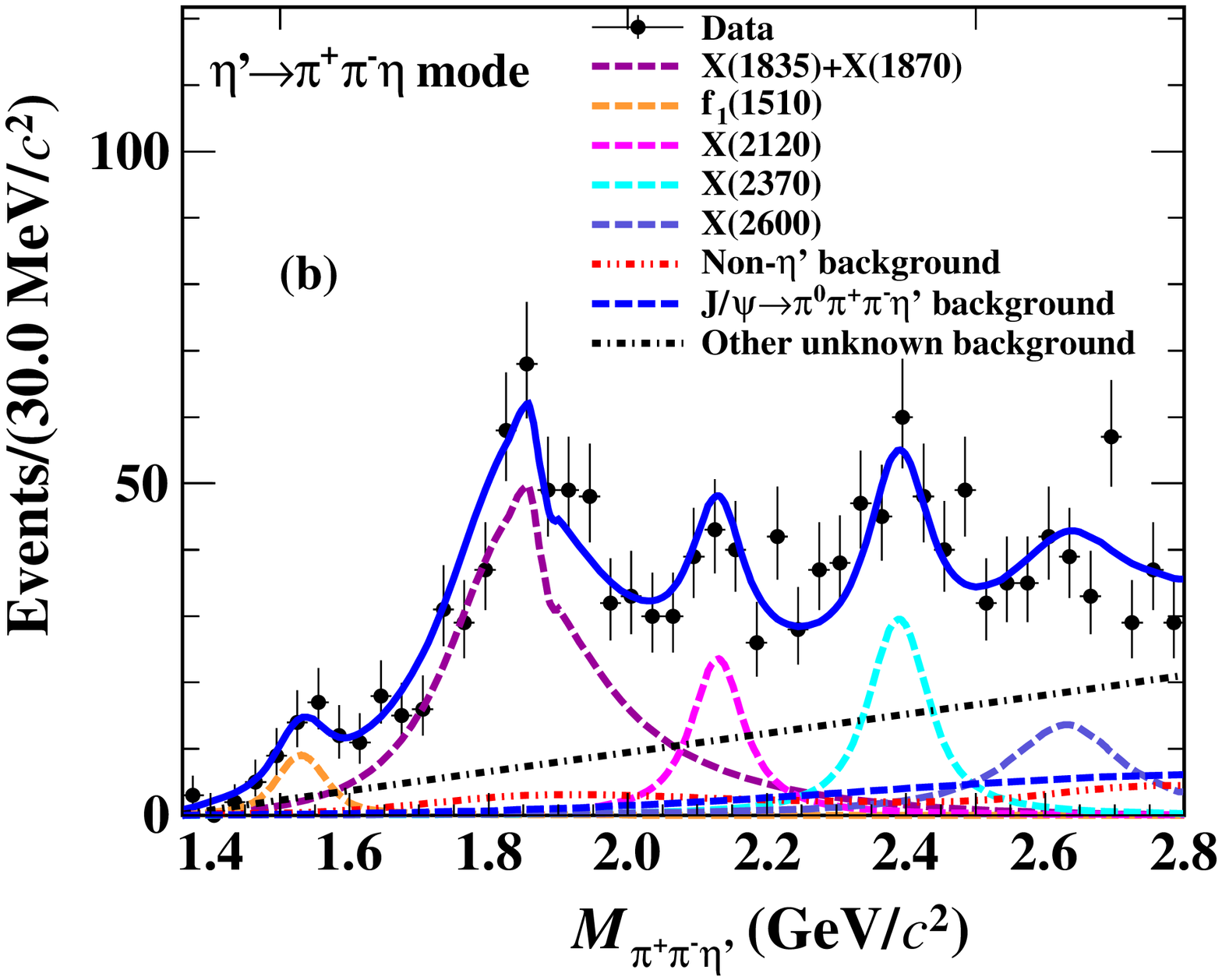}
  \caption{\small
Invariant mass distributions $M_{\pi^+\pi^-\eta'}$ for selected $\eta'\to\gamma\pi^+\pi^-$ candidates for the (a) $\eta'\to\gamma\pi^+\pi^-$ decay mode and (b) $\eta'\to\pi^+\pi^-\eta$ decay mode. The simultaneous fit results using a  sum of six Breit-Wigner amplitudes are overlaid. The dots with error bars are data, the blue solid line is the total fit, the orange dashed line describes the $f_1(1510)$ resonance, the purple dashed line -- the $X(1835)$ and $X(1870)$ states, the pink dashed line -- the $X(2120)$ state, the cyan dashed line -- the $X(2370)$ state, the light blue dashed line -- the $X(2600)$ state, the red dashed line indicates the non-$\eta'$ background, the blue dashed line is the $J/\psi\to\pi^0\pi^+\pi^-\eta'$ background, and the black dot-dashed line corresponds to a Chebyshev polynomial function describing other unknown backgrounds. 
}
\label{fig:fit}
\end{figure*}

After imposing the selection criteria described above, a clear $X(1835)$ signal is seen in the invariant mass spectra of the $\pi^+\pi^-\eta'$ candidates as shown in Fig.~\ref{fig:fit} for both $\eta'$ decay modes. 
Detailed event type analysis of the inclusive MC sample with a generic tool, TopoAna~\cite{topo}, shows that two classes of potential backgrounds are left after the candidates selection: events with no real $\eta'$ meson in the final states (non-$\eta'$), and events from $J/\psi\to\pi^0\pi^+\pi^-\eta'$ decays. 

In Ref.~\cite{lineshape}, two models were used to characterize the $\pi^+\pi^-\eta'$ line shape around 1.85 GeV/$c^2$: one incorporates the opening of a threshold in the mass spectrum (Flatt\'e formula), and the other is the coherent sum of two resonant amplitudes. Since our fit range reaches up to 2.8 GeV/$c^2$, the long and high tail of the Flatt\'e formula in the higher range of the first model makes the fit unstable. We therefore consider the other model which assumes the existence of a narrow resonance $X(1870)$ near the $p\bar{p}$ threshold, with the interference between this resonance and the $X(1835)$ causing the line-shape distortion. 
Thus, the anomalous line shape near 1.835 GeV/$c^2$ rate is modeled as $|T|^2$, where $T$ is the decay amplitude of a coherent sum of the two Breit-Wigner functions for $X(1835)$ and $X(1870)$, defined in Eq.~(3) of Ref.~\cite{lineshape}.  The parameters of $T$ in our fit are fixed to the measured values in Ref.~\cite{lineshape}. It is known that there should be two nontrivial solutions (constructive or destructive interference) in a fit using the coherent sum of two Breit-Wigner functions~\cite{twosolution}. The two solutions in our fit share the same mass and width of the $X(1835)$ and $X(1870)$ but the relative $\pi^{+} \pi^{-}\eta^{\prime}$ coupling strengths and the phase between them are different, leading to  different branching fractions of the two resonances. 

To determine the signal yields of the intermediate resonances in the $J/\psi\to e^+e^-\pi^+\pi^-\eta'$ decays, an simultaneous unbinned maximum likelihood fit is performed to the $M_{\pi^+\pi^-\eta'}$ spectra of selected data from 1.36 GeV/$c^2$ to 2.80 GeV/$c^2$ for the two $\eta'$ decay modes as shown in Fig.~\ref{fig:fit}. Due to low statistics and unknown variations across the $X$ Dalitz plot, the only interference considered is that within the amplitude $T$.  
The effect of the six intermediate resonances are described by the sum of five terms.  The first contains two resonances, 
\begin{equation}\label{eq2}
(|T|^{2} \times \epsilon_{\text {sig }}) \otimes g
\end{equation} 
and the remaining four terms are of the form
\begin{equation}\label{eq3}
(|BW|^{2} \times \epsilon_{\text {sig }}) \otimes g.
\end{equation}
Here, $\epsilon_{\text {sig }}$ is the mass-dependent detection efficiency; $g$ is a Gaussian function used to account for the mass resolution. 
The motivation for the $T$ in Eq.~\eqref{eq2} was discussed above.  
The other resonances, i.e. $f_1(1510)$, $X(2120)$, $X(2370)$ and $X(2600)$ are described by Eq.~\eqref{eq3} in the fit, where $|BW|^{2}$ is the Breit-Wigner function for $f_1(1510)$, $X(2120)$, $X(2370)$ and $X(2600)$. The masses and widths of the first three resonances are fixed to the values in Ref.~\cite{X2120}. 
Possible background contribution from non-$\eta'$ background processes is estimated by the events in the $\eta'$ mass sideband.  
To estimate the contribution from the $J/\psi\to\pi^0\pi^+\pi^-\eta'$ process, a phase space MC-simulation sample is generated and parametrized by a fourth order Chebyshev polynomial function. In addition to these two background processes, a first order Chebyshev polynomial function is used to describe other possible unknown background processes. The fit yields are $77\pm 23$, $1364\pm 53$, $310\pm 32$, $397\pm 37$ and $323\pm 62$ events for $f_1(1510)$, $X(1835)$ with $X(1870)$, $X(2120)$, $X(2370)$ and $X(2600)$, respectively. The statistical significance is determined from the change of the log likelihoods with and without the corresponding signal shape. 
The effect of the assumed background and signal shapes are also considered when evaluating the significance.  Variations are tested, and the lowest significances of the $f_1(1510)$, $X(1835)$, $X(2120)$, $X(2370)$ and $X(2600)$ states are calculated to be 3.3$\sigma$, 15$\sigma$, 5.3$\sigma$, 7.3$\sigma$ and 3.2$\sigma$, respectively. 

\begin{figure}[htp]
  \centering
\includegraphics[width=0.8\linewidth]{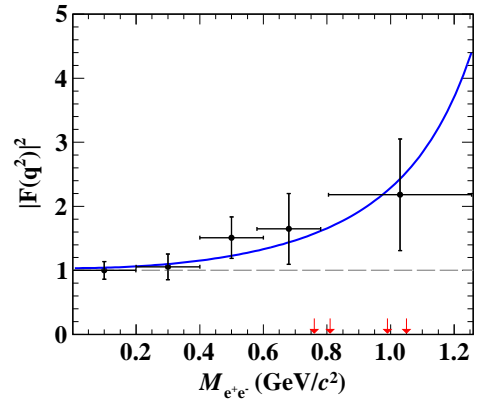}
  \caption{\small
$|F(q^2)|^{2}$ distribution for $J/\psi\to e^+e^-X(1835)$ decays. The dots with error bars are $|F(q^2)|^{2}$ values, the solid blue curve is the fit result according to the simple pole approximation and the gray dashed line represents $|F(q^2)|^{2}=1$. The red arrows denote the $M_{e^+e^-}$ veto-requirements.}
\label{fig:TTF}
\end{figure}

The TFF $|F(q^2)|^2$ for the process $J/\psi\to e^+e^-X(1835)$ is measured by dividing the $M_{e^+e^-}$ distribution into five intervals. The values of $|F(q^2)|^2$s are obtained as the ratios of branching fractions measured with the fit method described above and those predicted by QED in each interval as shown in Eq.~\eqref{eq0}. 
These QED predicted branching fractions are obtained from Eq.~(12) of Ref.~\cite{formfactor}. Figure~\ref{fig:TTF} shows the values of $|F(q^2)|^2$ determined in each interval of the  $M_{e^+e^-}$ distribution. A simple pole approximation parameterized as $F(q^2)=\frac{1}{1-q^{2} / \Lambda^{2}}$~\cite{formfactor, DIY}, where the parameter $\Lambda$ is the spectroscopic pole mass, is used to fit the distribution. From the fit, $\Lambda$ is determined to be ($1.75\pm 0.29(\mathrm{stat})\pm 0.05(\mathrm{syst})$) GeV/$c^2$. The systematic uncertainty is taken as the difference between the baseline $\Lambda$ value and the refitted one including the systematic uncertainties of the measured branching fractions in each interval.

The systematic uncertainties on the branching fractions associated with the efficiency include the signal generator, photon detection efficiency, tracking efficiency, PID efficiency, kinematic fit, mass spectra requirements ($\gamma e^+e^-$, $\pi^+\pi^-$, $\pi^+\pi^-\gamma(\eta)$ and $e^+e^-$) and $R_{xy}$ requirement. The uncertainties from these efficiency-related sources are 3.7\%, 4.9\% and 7.2\% for the $X(1835)$, $X(2120)$ and $X(2370)$, respectively.
The total systematic uncertainties related to fit methods include the non-$\eta'$ background, $J/\psi\to\pi^0\pi^+\pi^-\eta'$ background, unknown background and signal shape. 
The total uncertainties due to the fit are 1.3\%, 4.5\% and 4.7\% for the $X(1835)$, $X(2120)$ and $X(2370)$, respectively.  Additional uncertainties for the total number of $J/\psi$ events~\cite{njpsi2017}, and the $\eta'$ branching fractions to the final states $\gamma\pi^+\pi^-$ and $\pi^+\pi^-\eta$~\cite{pdg2020} are also included. The total for all systematic uncertainties on the product branching fractions of the $X(1835)$, $X(2120)$ and $X(2370)$ modes are 4.3\%, 6.8\% and 8.8\%, respectively. 
The final branching fraction results are shown in Table~\ref{tab:br}.

\begin{table}[htp]
\centering
\caption{\label{tab:br}\small 
 Branching fractions of $J/\psi\to e^+e^-X$, $X\to\pi^+\pi^-\eta'$ decays. The first uncertainties are statistical, and the second are systematic.}
 \begin{tabular}{ll}
\hline \hline \multicolumn{2}{c} {Branching fractions of $J/\psi\to e^+e^-X$, $X\to\pi^+\pi^-\eta'$} \\\hline
$X=X(1835)$ (solution I)& $\left(3.58 \pm 0.19 \pm 0.16\right) \times 10^{-6}$ \\
 ~~~~~~~~~~~~~~~~~~ (solution II)& $\left(4.43 \pm 0.23 \pm 0.19\right) \times 10^{-6}$ \\\hline
$X=X(2120)$&$\left(0.82 \pm 0.12 \pm 0.06\right) \times 10^{-6}$\\\hline
$X=X(2370)$&$\left(1.08 \pm 0.14 \pm 0.10\right) \times 10^{-6}$\\
\hline \hline
\end{tabular}
\end{table}

In summary, using a sample of about 10 billion $J/\psi$ events collected at the center-of mass energy $\sqrt s = 3.097$ GeV with the BESIII detector, we report the observation of the EM Dalitz decay $J/\psi \to e^+e^- \pi^+ \pi^- \eta'$. This is also the first observation of the states $X(1835)$, $X(2120)$ and $X(2370)$ in the EM Dalitz decays, and the first measurement of the TFF between $J/\psi$ and $X(1835)$. 
According to the model of a coherent sum of two Breit-Wigner amplitudes of $X(1835)$ and $X(1870)$, the branching fraction of $J/\psi\to e^+e^-X(1835)$, $X(1835)\to\pi^+\pi^-\eta'$ is measured to be $(3.58\,\pm\,0.19(\mathrm{stat})\,\pm\,0.16(\mathrm{syst})) \times 10^{-6}$ (constructive interference)/ $(4.43\,\pm\,0.23(\mathrm{stat})\,\pm\,0.19(\mathrm{syst})) \times 10^{-6}$ (destructive interference) with a significance of $15\sigma$. With respect to the radiative decay $J / \psi \rightarrow \gamma X(1835),~ X(1835)\rightarrow \pi^+\pi^-\eta'$ \cite{lineshape}, the ratio $R$ of the branching fractions is determined to be $(1.19\,\pm\,0.10(\mathrm{stat})\,\pm\,0.14(\mathrm{syst}))$$\times10^{-2}$. The measured $R$ is consistent with the theoretical prediction~\cite{ratio} within two standard deviations ($2\sigma$). The branching fractions of $J/\psi\to e^+e^-X(2120),~ X(2120)\to\pi^+\pi^-\eta'$ and $J/\psi\to e^+e^-X(2370),~X(2370)\to\pi^+\pi^-\eta'$ are measured to be $(0.82\,\pm\,0.12(\mathrm{stat})\,\pm\,0.06(\mathrm{syst}))\times10^{-6}$ and $(1.08\,\pm\,0.14(\mathrm{stat})\,\pm\,0.10(\mathrm{syst})) \times 10^{-6}$ with significances of 5.3$\sigma$ and 7.3$\sigma$, respectively. The measured values of $|F(q^2)|^2$ for the $J / \psi \rightarrow e^{+} e^{-} X(1835)$ channel deviate from the point-like particle assumption ($|F(q^2)|^2 = 1$) significantly and have been parametrized in the simple pole approximation as $F(q^2)=\frac{1}{1-q^{2} / \Lambda^{2}}$ with $\Lambda = (1.75\,\pm\,0.29(\mathrm{stat})\,\pm\,0.05(\mathrm{syst})$) GeV/$c^2$. This measured pole mass $\Lambda$ can be used as an input parameter to improve the theoretical calculations. 

The BESIII collaboration thanks the staff of BEPCII and the IHEP computing center for their strong support. The authors thank Xian-Wei Kang for useful discussions. This work is supported in part by National Key R\&D Program of China under Contracts Nos. 2020YFA0406300, 2020YFA0406400; National Natural Science Foundation of China (NSFC) under Contracts Nos. 11805037, 11625523, 11635010, 11735014, 11822506, 11835012, 11935015, 11935016, 11935018, 11961141012, 12022510, 12025502, 12035009, 12035013, 12061131003; the Chinese Academy of Sciences (CAS) Large-Scale Scientific Facility Program; Joint Large-Scale Scientific Facility Funds of the NSFC and CAS under Contracts Nos. U1832121, U1732263, U1832207; CAS Key Research Program of Frontier Sciences under Contract No. QYZDJ-SSW-SLH040; 100 Talents Program of CAS; INPAC and Shanghai Key Laboratory for Particle Physics and Cosmology; ERC under Contract No. 758462; European Union Horizon 2020 research and innovation programme under Contract No. Marie Sklodowska-Curie grant agreement No 894790; German Research Foundation DFG under Contracts Nos. 443159800, Collaborative Research Center CRC 1044, FOR 2359, GRK 214; Istituto Nazionale di Fisica Nucleare, Italy; Ministry of Development of Turkey under Contract No. DPT2006K-120470; National Science and Technology fund; Olle Engkvist Foundation under Contract No. 200-0605; STFC (United Kingdom); The Knut and Alice Wallenberg Foundation (Sweden) under Contract No. 2016.0157; The Royal Society, UK under Contracts Nos. DH140054, DH160214; The Swedish Research Council; U. S. Department of Energy under Contracts Nos. DE-FG02-05ER41374, DE-SC-0012069.

\nolinenumbers

\onecolumngrid

\end{document}